\documentclass[11pt,a4paper]{article}
\pdfoutput=1
\usepackage{jheppub}
 \newcommand{\PRE}[1]{}
\usepackage{amsmath,amsfonts,epsfig,graphicx}
\usepackage{multirow}
\usepackage{indentfirst}
\usepackage[latin1]{inputenc}
\usepackage{amssymb}
\usepackage{verbatim}
\usepackage{float}
\usepackage{subfig}
\usepackage{slashed} 
\usepackage{siunitx}
\usepackage{hyperref}
\usepackage{todonotes}

\newcommand{\al}{{\it et al. }}
\newcommand{\MG}{M{\scriptsize AD}G{\scriptsize RAPH}5\_{\scriptsize aMC@NLO}}

\begin{document}

\title{Searching for heavy leptoquarks at a muon collider}

\author[a]{Sitian Qian,}
\author[a]{Congqiao Li,}
\author[a]{Qiang Li,}
\author[a]{Fanqiang Meng,}
\author[a]{Jie Xiao,}
\author[a]{Tianyi Yang,}

\author[b]{Meng Lu,}
\author[b]{and Zhengyun You}

\affiliation[a]{
Department of Physics and State Key Laboratory of Nuclear Physics and Technology, Peking University, Beijing, 100871, China}
\affiliation[b]{
School of Physics, Sun Yat-Sen University, Guangzhou 510275, China}
\emailAdd{sitian.qian@cern.ch}
\emailAdd{meng.lu@cern.ch}
\emailAdd{qliphy0@pku.edu.cn}

\abstract{
The LHCb Collaboration recently gave an update on testing lepton flavour universality with $B^+ \rightarrow K^+ \ell^+ \ell^-$, in which a 3.1 standard deviations from the standard model prediction was observed. The g-2 experiment also reports a 3.3 standard deviations from the standard model on muon anomalous magnetic moment measurement. These deviations could be explained by introducing new particles including leptoquarks. In this paper, we show the possibility to search for heavy spin-1 leptoquarks at a future TeV scale muon collider by performing studies from three channels:  1) same flavour final states with either two bottom or two light quarks, 2) different flavour quark final states, and 3) a so-called ``VXS" process representing the scattering between a vector boson and a leptoquark to probe the coupling between leptoquark and tau lepton. We conclude that a 3 TeV muon collider with $3~\si{ab^{-1}}$ of integrated luminosity is already sufficient to cover the leptoquark parameter space in order to explain the LHCb lepton flavour universality anomaly.  
}

\maketitle


\section{Introduction}

Lepton Flavour Universality (LFU), one of the fundamental assumptions of the Standard Model (SM), states that the lepton family shares common properties when interacting with gauge bosons regardless of their different masses. However, there have been clues that LFU seems to be broken and thus leads to the evidence of physics beyond the SM (BSM). One interesting result is that the Muon g-2 experiment announces very recently that muon anomalous magnetic moment has 3.3 standard deviations from SM~\cite{gminus2}, which is consistent with the previous measurement at BNL~\cite{gminus2-2} and leads to a combined 4.2 standard deviations from SM. Another interesting measurement is performed on the $R_K$ anomaly through rare $B$ meson decay from LHCb  Collaboration~\cite{Rk1}, where the $R_K$ is defined as the ratio of branching fractions:
\begin{equation}
    R_K=\frac{BR(B^+\rightarrow K^+\mu^+\mu^+)}{BR(B^+\rightarrow K^+e^+e^+)},
\end{equation}
the results show an evidence of breaking of LFU at 3.1 standard deviations from the SM.
\begin{figure}[htbp]
    \centering
    \subfloat[Muon $g-2$ experiment]{\includegraphics[width=0.4\textwidth]{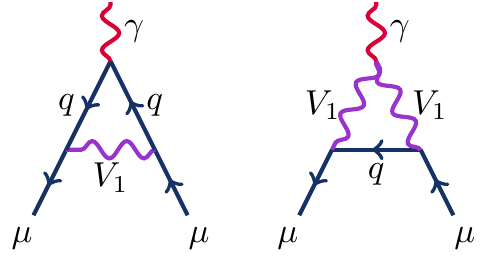}}\qquad
    \subfloat[LHCb B anomaly]{\includegraphics[width=0.4\textwidth]{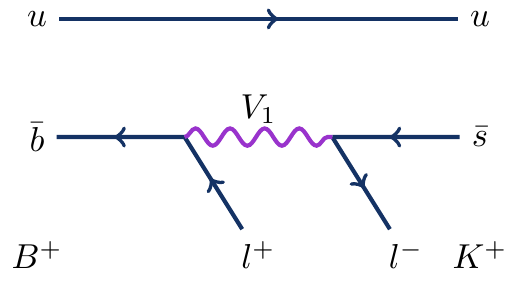}}
    \caption{LQ induced Feynman Diagrams explaining both muon anomalies}
    \label{fig:B_anomaly_FD}
\end{figure}

Discrepancy between SM prediction and measurement of muon anomalous magnetic moment is against the expectation of physicists, as for the theoretical prediction  and measurement of electron magnetic moment are consistent with each other at a precision of a part per billion~\cite{gminus2-3}, which is one of the most powerful evidence of the SM. In the other hand, there have been several previous measurements on $R_K$ from LHCb Collaboration~\cite{Rk2,Rk3}, the continuous measurements and consistent results draw interests on the LFU anomaly gradually. Due to the suppress of flavour-changing neutral current, this decay channel of $B$ meson only occurs at loop level and supposed to be very rare~\cite{rare_bdecay}, which makes this channel sensitive to the potential deviation or anomaly.

An interesting and straightforward interpretation for both the muon magnetic and LFU anomalies is to introduce the leptoquark (LQ)~\cite{LQmodel0,LQmodel1,LQmodel2,LQmodel3,LQmodel4,LQmodel5,LQmodel6,LQmodel7,LQmodel8,LQmodel9,LQmodel10,LQmodel11,LQmodel12,LQmodel13,LQmodel14,LQmodel15,LQmodel16,LQmodel17} which couples to a lepton and a quark at the same time. Figure.~\ref{fig:B_anomaly_FD} shows the typical Feynman diagrams explaining how the LQ contributions accommodate both two anomalies. Contribution to the muon anomalous magnetic moment from LQ is discussed in detail in Ref.~\cite{muong2_1,muong2_2,muong2_3,muong2_4,muong2_5,muong2_6,muong2_7,muong2_8,muong2_9,muong2_10}. The SM particle zoo welcomes LQs of different forms, such as scalar triplet, vector singlet and vector triplet. Despite of the representation, LQs interact with charged leptons and down-type quarks and the strength can be described phenomenologically by the coupling matrix:
\begin{equation}
{\lambda_{\overline{Q}L}} =  
\begin{pmatrix} 
\lambda_{de}&\lambda_{d\mu}&\lambda_{d\tau}\\
\lambda_{se}&\lambda_{s\mu}&\lambda_{s\tau}\\
\lambda_{be}&\lambda_{b\mu}&\lambda_{b\tau} 
\end{pmatrix}.
\vspace{0.5cm}
\label{eq:coupling_full}
\end{equation}

While LQ from each kind of representation can explain the muon anomalies in its own way, in this paper we will only introduce vector singlet LQs as the complementation to SM, whose Lagrangian is formulated  as Eq.~(\ref{eq:lq_lag}) by following~\cite{LQ1}: 
\begin{equation}
    \mathcal{L}_{\text{int}} = \left(\lambda_{\bar{Q}L}\bar{Q}_L\gamma_\mu L_L+\lambda_{\bar{D}E}\bar{D}_R\gamma_\mu E_R\right)V_1^\mu + \text{h.c.}
    \label{eq:lq_lag}
\end{equation}
here $Q_L$($L_L$) stands for left-handed SM quark(lepton) doublet, and $D_R$($E_R$) stands for the right-handed SM down-type quark(charged lepton) as the consequence of the absence of SM right-handed neutrino.

To explain the $R_K$ anomaly in the context of leptoquark models, one can relate $R_K$ to LQs coupling coefficients $\lambda_{\bar{Q}L}$. Constraint from $R_K$ and $R_K^*$ reads~\cite{LQ1}
\begin{equation}
    \frac{\lambda_{b\mu}\bar\lambda_{s\mu} - \lambda_{be}\bar\lambda_{se}}{M_{\text{LQ}}^2} \simeq -\frac{1\pm0.24}{(40~\si{TeV})^2}.
    \label{eq:lqconstrain}
\end{equation}
Despite that the LQ coupling with electron is also able to explain for the $R_K$ anomaly, anomaly from muon flavour is more favoured by other experimental results, for example the anomalous magnetic moment measurement~\cite{gminus2-3} and $b\to s\mu\mu$ angular distributions~\cite{LQ1}. In this paper we will only focus on coupling between LQ and $\tau$ or $\mu$ leptons. More specifically, we are interested in the coupling matrix with the form

\begin{equation}
{\lambda_{\overline{Q}L}} =  
\begin{pmatrix} 0&0&0\\0&\lambda_{s\mu}&\lambda_{s\tau}\\0&\lambda_{b\mu}&\lambda_{b\tau} \end{pmatrix},
\vspace{0.5cm}
\label{eq:coupling}
\end{equation}
where we assume couplings between LQ and the first generation lepton is zero. Despite that the coupling for first generation quarks are set to be zero nominally, LQs are still able to talk with $d$ quarks in the context of Cabibbo-Kobayashi-Maskawa (CKM) mixing. An additional assumption of this paper is that each coupling constant $\lambda_{\bar{Q}L}$ is treated as a real number.

There are quite a lot proposals for the next generation collider for the purpose of Higgs boson related measurements, among which the lepton colliders are in the majority. The promising proposals include Linear or Circular electron positron collider~\cite{ILC, FCC, CEPC, CLIC} and muon collider~\cite{Muc0,Muc1,Muc2,Muc3,Muc4}. Compared with the existing hadron colliders, muons collider could be more powerful with cleaner environment. Muons can be accelerated to much higher energy than electrons due to much less synchrotron radiation in a circular accelerator. These advantages of muon collider and recent muon anomalies, in addition to the assumption that LQs do not couple to electrons, drive us to perform the LQs study at a muon collider. There are already some interesting results of LQs searches at muon collider~\cite{LQ_RK1,LQ_RK2,LQ_RK3}, our study will be supplement to them and make it clear that the muon collider is able to answer the question of $B$ meson rare decay anomaly. It is a well-known fact that the challenge of a muon collider is to provide high luminosity and high phase-space density muon beams due to the short life time and the source of muons (produced from $\pi$-decay)~\cite{Mu1,Mu2}. Furthermore, the  beam-induced background (BIB)~\cite{BIB1,BIB2,BIB3,BIB4,BIB5,BIB6,BIB7,BIB8} from muon decays is also tricky to handle. However, we assume here the advanced technique has been achieved to provide high luminosity muon beams with high quality in the future muon collider, and the BIB issue is also solved with advanced detector design (e.g., using timing detector or a shielding nozzle). Our studies will be shown based on these assumptions. We also choose moderate work condition of a benchmark muon collider by choosing center-of-mass (C.O.M) energy $\sqrt{s}=3~\si{TeV}$ with integrated luminosity $L=3~\si{ab ^{-1}}$. 

\section{Leptoquark Search at a Muon Collider}
\subsection{Samples and cuts}
The samples we used are generated using \MG (MG5aMC)~\cite{MG5}. The BSM leptoquark model is implemented by Ref.~\cite{LQ1} with FeynRules~\cite{feynrules} in the Universal FeynRules Output convention.

In MG5aMC, basic cuts on phase space are applied as Table~\ref{tab:Cuts} documented. Cuts on light jet and charged leptons follows the default setting of MG5aMC. The transverse momentum requirement of b jet is aligned with the light jet case, while the pseudo-rapidity cut on b jet is aligned with the charged lepton as they both need track information provided by the tracker detector.

\begin{table}[htbp]
    \centering
    \begin{tabular}{|c|c|c|c|}
    \hline
    Category & light jet & b jet & charged lepton \\
    \hline
      $p_T$   &  $> 20$ GeV & $> 20$ GeV & $> 10$ GeV\\
      \hline
        $|\eta|$ & $< 5$ & $< 2.5$ & $< 2.5$ \\
        \hline
    \end{tabular}
    \caption{Phase space cuts used in MG5aMC calculations.}
    \label{tab:Cuts}
\end{table}

\subsection{Di-jet Final State}
\begin{figure}[htbp]
    \centering
    \subfloat[Different Flavour $t$-channel Di-jet Final State\label{sfig:DF_Diagram}]{\includegraphics[width=0.25\textwidth]{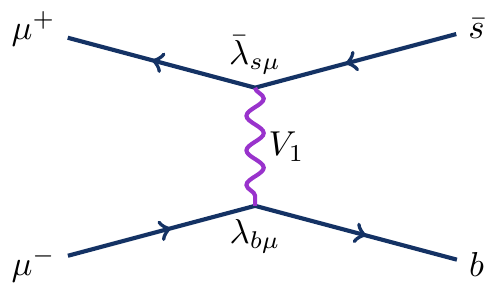}}\qquad
    \subfloat[Same Flavour $t$-channel Di-jet Final State\label{sfig:SF_Diagram}]{\includegraphics[width=0.25\textwidth]{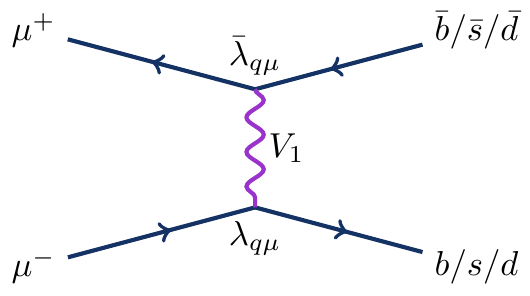}}\qquad
    \subfloat[SM $s$-channel Di-jet Production\label{sfig:DY_Diagram}]{\includegraphics[width=0.25\textwidth]{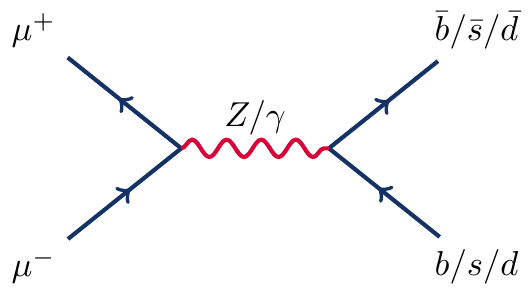}}
    \caption{Di-jet production mechanism at a muon collider with Leptoquarks.}
    \label{fig:MuC_Dijet}
\end{figure}

At a muon collider, the simplest processes involving LQs would be the di-jet final states in which LQs show up in $t$-channel and lead to the muon-quark conversion of incoming muons (Fig.~\ref{sfig:DF_Diagram}~\&~\ref{sfig:SF_Diagram}). The di-jet in the final states can be either different flavour or same flavour, while the later one will interfere with the SM $s$-channel di-jet production (Fig.~\ref{sfig:DY_Diagram}).

Those $t$-channel processes can be classified into three categories as shown in Table~\ref{tab:Dijet_Category}, where ``light (jet)'' stands for jet originating from Matrix Element (ME) level $u/d/c/s/g$ partons. The cross section dependence on the BSM coupling strength of different flavour final state is simple since there is no such SM contribution, as shown in Table.~\ref{tab:Dijet_Category}. The situation for same flavour final state is complicated due to the interference between BSM signal and SM $s$-channel process, as shown in Fig.~\ref{sfig:Xsec_Comp}, and we can also find the obvious destructive interference between SM $s$-channel background and LQ $t$channel signal.

\begin{table*}[htbp]
    \centering
    \begin{tabular}{|c|c|c|}
    \hline
       Category  & SM $s$-channel Interference & Cross Section Dependence  \\
    \hline
        light+$b$ & No & $|\lambda_{s\mu}\bar{\lambda}^{b\mu}|^2$ \\
        \hline
        light+light & Yes & Complicated \\
        \hline
        $b$ + $b$ & Yes & Complicated \\
        \hline
    \end{tabular}
    \caption{LQ search strategy through the  di-jet processes.}
    \label{tab:Dijet_Category}
\end{table*}

\subsubsection{Same Flavour Final State}

To determine the LQ coupling strengths exclusively, we start with the same flavour final states. In Ref.~\cite{LQ_RK2} the authors discuss same flavour dijet state in detail however mainly focusing on $b+b$ flavour. As the $B$ meson rare decay anomaly requires non-zero product between the quark-LQ-muon coupling constants of $b$ and $s$ flavour, specialized flavour assumptions are needed to translate di-bjet limits to di-light jet limit in that case. In our study, we will perform analysis on both $b+b$ and light+light final states thus leads to a direct comparison between our results and the B meson anomaly measurement.

\begin{figure}[htbp]
    \centering
    \subfloat[Cross section dependence on $|\lambda_{b\mu}|$ for a 1 and 10 TeV LQ, respectively.~\label{sfig:Xsec_Comp}]{\includegraphics[width=0.4\textwidth]{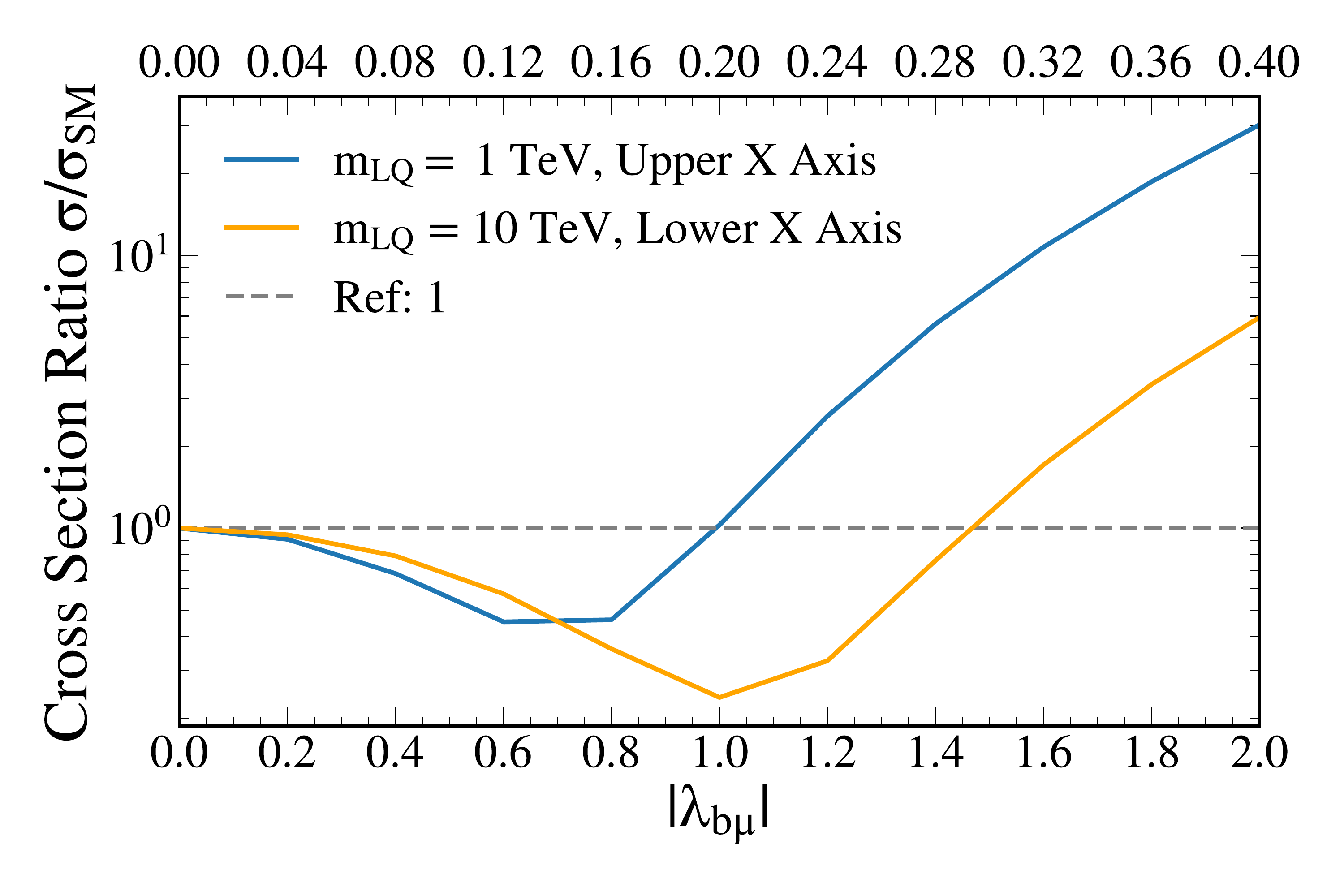}}\qquad
    \subfloat[Shape comparison on $|\eta_{b}|$ distribution for the SM, and the LQ cases with $|\lambda_{b\mu}|=2$\label{sfig:Shape_Comp}]{\includegraphics[width=0.4\textwidth]{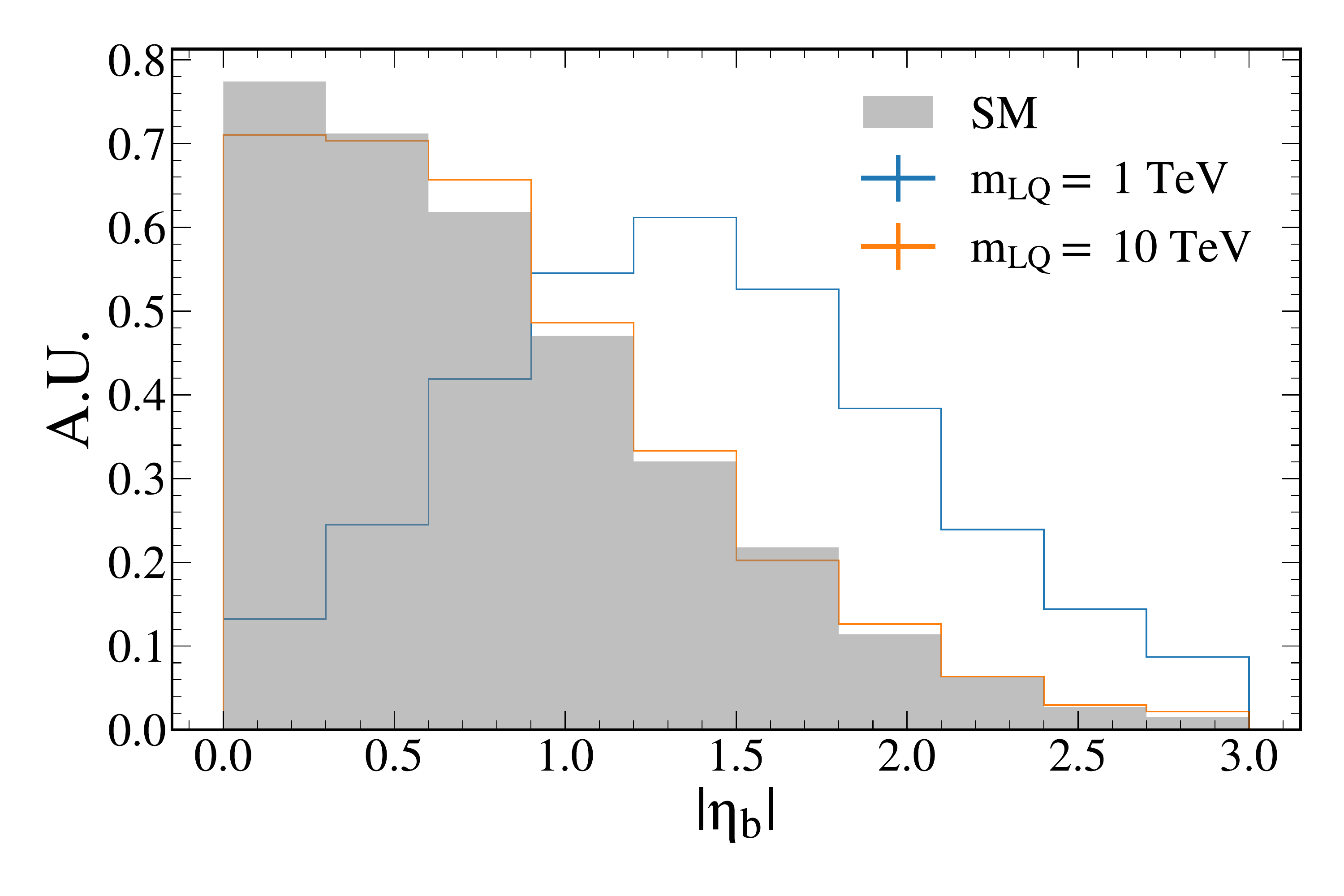}}
    \caption{Effects of a 1 TeV and 10 TeV leptoquark in the $b+b$ di-jet process.}
    \label{fig:Compare}
\end{figure}

The same flavour di-jet events are generated considering the interference between BSM signal and SM $s$-channel background. In Fig.~\ref{fig:Compare} we use the $b+b$ final state for illustration while the light+light final state shares a similar behaviour. Please note that the rapid growing of cross section is caused by so-called unitarity violation (UV), which is a compromise on allowing LQs couple to all lepton generations as discussed in Ref.~\cite{LQ1}. Figure.~\ref{sfig:Shape_Comp} shows the shape comparison of $|\eta_b|$ distribution with $|\lambda_{b\mu}|=2$, while all the shapes are normalized to cross section of SM value. The shape of $|\eta_b|$ is used for statistic analysis and ten bins are used in the range of (0, 2.5), while the yields are estimated using the corresponding cross section at 3 TeV C.O.M energy with integrated luminosity $3~\si{ab^{-1}}$.

We apply the same analysis strategy for both $b+b$ final state and light+light final states. Apart from the $s+s$ final state events, s+d and d+d final states can still appear after setting $\lambda_{d\mu}$ to zero due to CKM mixing. However, as our light jet definition includes all $u/d/s/c$ quarks, the total expected yields of light+light final state will only depend on the $\lambda_{s\mu}$ deterministically and thus can be handled consistently once the strength of CKM mixing is fixed. In our study, we assume the LQs couple with the SM weak eigenstate of down-type quarks, thus the CKM mixing is set to the SM Cabibbo angle.

\begin{figure}
    \centering
    \subfloat[$b+b$ final state\label{sfig:bb_Res}]{\includegraphics[width=0.4\textwidth]{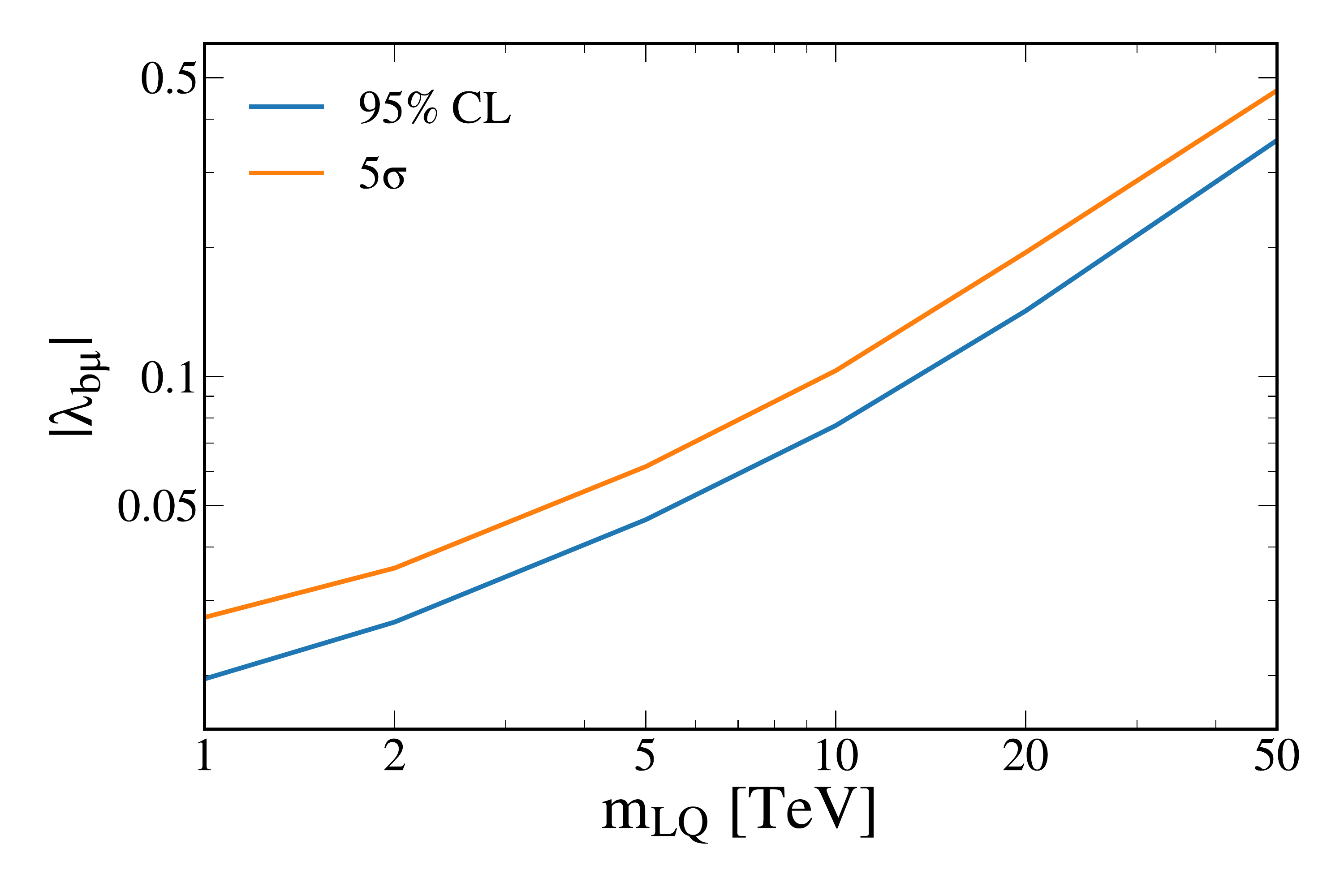}
    }\qquad
    \subfloat[light + light final state\label{sfig:ll_Res}]{\includegraphics[width=0.4\textwidth]{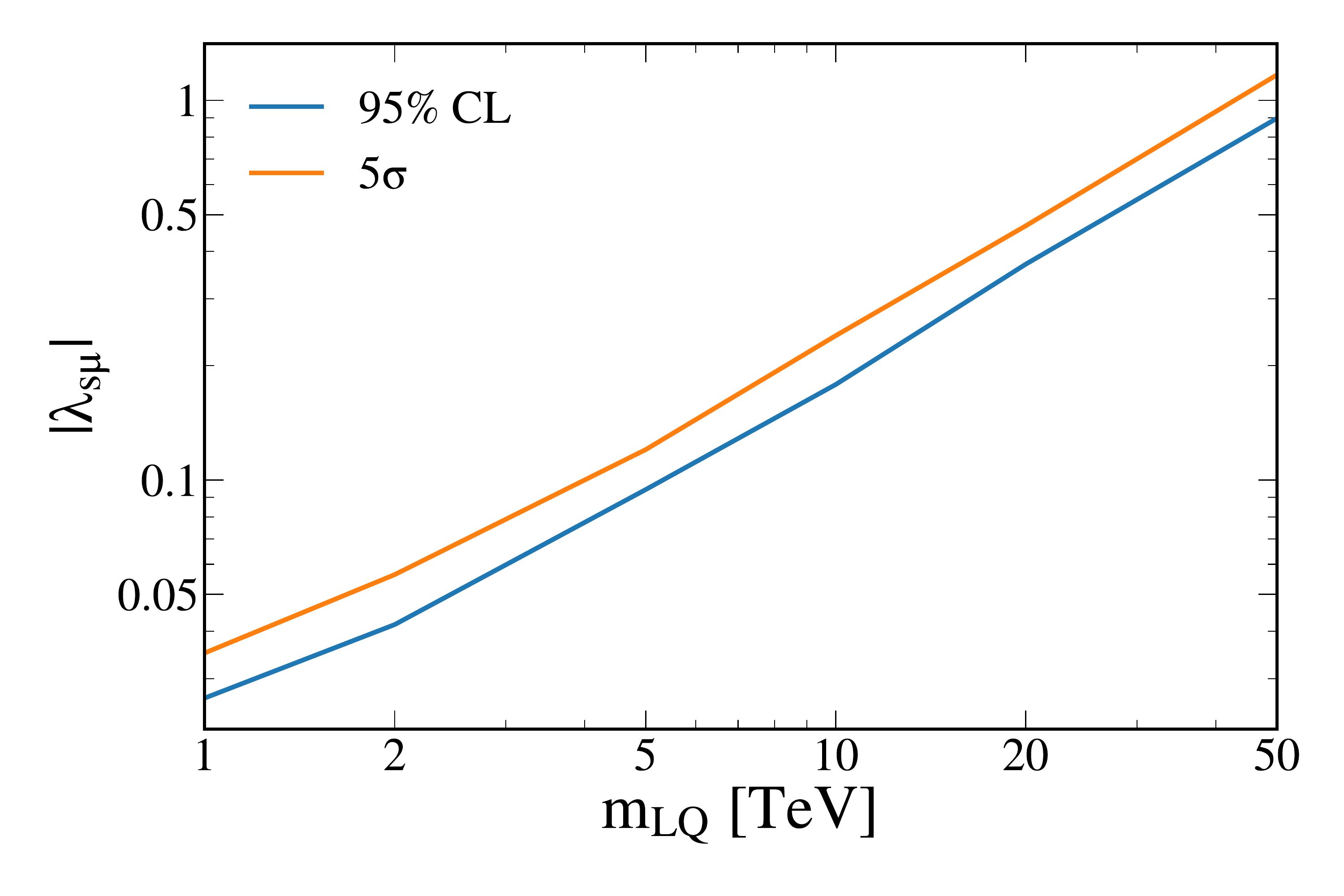}}
    \caption{95\% CL exclusion line (dashed) and 5$\sigma$ discovery limit (solid) against different vector singlet LQ mass values.}
    \label{fig:SF_Res}
\end{figure}

Though our analysis is performed at parton level, we introduce jet flavour identification efficiency mimicking the jet tagging techniques~\cite{btag} in CMS in order to obtain realistic results. The jet flavour tagging efficiencies we used are listed in Table~\ref{tab:jet_eff}. We should keep in mind that the benchmark efficiency we choose is at moderate transverse momentum of jets, and can be improved up to TeV scale using e.g. the machine learning technique~\cite{btag,btag_deep,btag_atlas,btag_mux} and may be further improved at the muon collider (one may also rely on di-jet + QED/QCD emission processes to lower down the momenta of final state jets). Currently the jet tagging techniques for muon colliders are in a preliminary stage~\cite{muon_btag_1,muon_btag_2,muon_btag_3} and the methodology on reconstructing the secondary vertex is the same as the hadron collider ones. Thus, hadron collider tagging efficiencies can be viewed as a solid baseline. Although there could be ``fake'' background from mis-tagged di-jet events, we neglect its contribution since it is negligible compared with the  dominant contribution from SM $s$-channel background. We use the statistics $Z$ defined in Eq.~\ref{eq:SF_Stats} for both 95\% CL exclusion limit and 5$\sigma$ discovery limit, where $b$ is the SM background, $n:=s+b$ is the total yields containing both signal and background, $s$ is the BSM signal. Both $Z$ statistics subject to $\chi^2$ distribution with 10 degree of freedom corresponding to 10 bins~\cite{asymptotic}. Results are summarized and plotted in Fig.~\ref{fig:SF_Res} for both final states. 

\begin{equation}
\begin{aligned}
    &Z = \sum_{i=1}^{10} Z_i \\
    &\begin{cases}
    Z_i := \left[ n_i - b_i + b_i\ln(b_i/n_i)\right]&\text{95\% CL Exclusion}\\
    Z_i := \left[ b_i - n_i + n_i\ln(n_i/b_i)\right]&\text{$5\sigma$ Discovery}
    \end{cases}
\end{aligned}
    \label{eq:SF_Stats}
\end{equation}

\begin{figure}
    \centering
    \subfloat[1 TeV LQ combined results]{\includegraphics[width=0.4\textwidth]{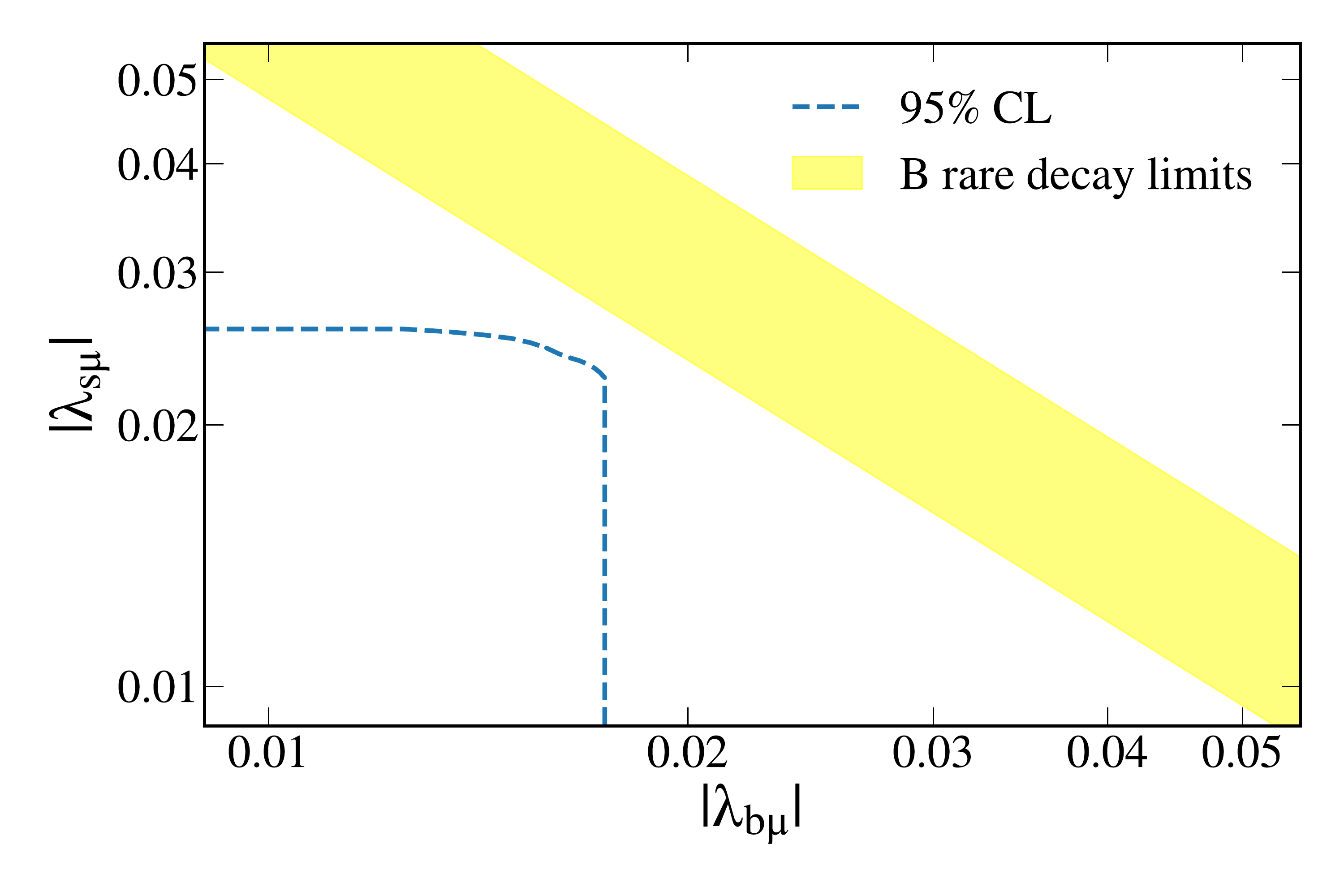}}\qquad
    \subfloat[10 TeV LQ combined results]{\includegraphics[width=0.4\textwidth]{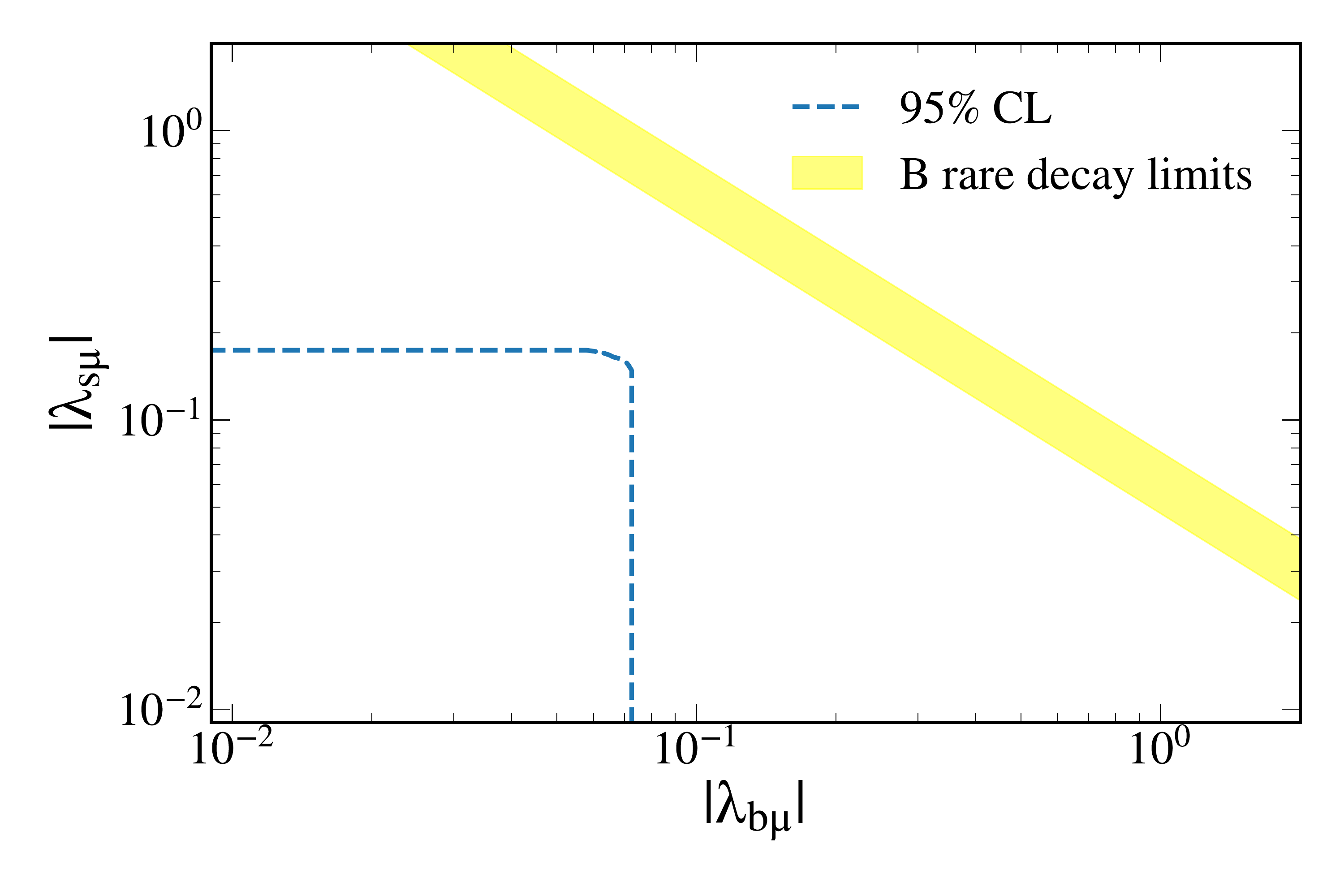}}
    \caption{95\% CL exclusion contour line on the $|\lambda_{b\mu}|-|\lambda_{s\mu}|$ plane. Upper right portion of the dashed blue line is the exclusion region.}
    \label{fig:Contour}
\end{figure}

Treating the $b+b$ and light $+$ light final state as different measurement and combining the p-values with the so-called Fisher method~\cite{fisher}, we can obtain the corresponding 2d 95\% CL exclusion contour as Fig.~\ref{fig:Contour} shows. Last but not least, the resulting limits on LQ couplings $\lambda_{ql}$ are relatively small compared with those values resulting in very high cross sections (which may be close to UV) in Figure~\ref{sfig:Xsec_Comp}, thus the results are safe even with a UV incomplete BSM LQ model.
\begin{table}[htbp]
    \centering
    \begin{tabular}{|c|c|c|}
    \hline
        Flavour & Tagged as light & Tagged as $b$ \\
        \hline
        light & 0.9 & 0.01 \\
        \hline
        $b$ & 0.1 & 0.7 \\
        \hline
    \end{tabular}
    \caption{Flavour tagging efficiencies used in this analysis.}
    \label{tab:jet_eff}
\end{table}

\subsubsection{Different Flavour Final State}

Among all possible $t$-channel di-jet states, the different flavour channel is an appealing choice for LQ search as there is no such process within SM due to the fact that processes with different flavour di-jet violate the conservation law on quark numbers. In Ref~\cite{LQ_RK1}, the authors discuss a similar final state via introducing a scalar leptoquark, however with $b+s$ final states only without concerning the potential CKM mixing. 

The ME cross section of such BSM process will proportional to the product of coupling constants associated with two muon-LQ-quark vertices as shown in Fig.~\ref{sfig:DF_Diagram}. The corresponding cross section reads
\begin{equation}
    \sigma_{\text{DF}}(\mu^+\mu^-\to jj) \propto |\lambda_{s\mu}\bar{\lambda}^{b\mu}|^2
    \label{eq:df_xsec}
\end{equation}
as described in Table~\ref{tab:Dijet_Category}, where DF means ``Different Flavour''. However, the different flavour channel is not background free as there can be ``fake'' contributions originating from jet mis-tag, e.g., one b jet is identified as a light jet in a di-bjet event or vice versa.

We set up the statistic analysis using event counting method to estimate the limits of coupling strengths at 95\% CL as well as the 5$\sigma$ discovery of BSM signal. For each coupling strength, we calculate the corresponding signal and background yields, and construct the corresponding $Z$ statistics for each case following Eq.~\ref{eq:DF_Stats}. The $Z$ statistic subjects to a $\chi^2$ distribution with 1 degree of freedom~\cite{asymptotic}.

\begin{equation}
    \begin{cases}
    Z := \left[ n - b + b\ln(b/n)\right]&\text{for 95\% CL Exclusion}\\
    Z := \left[ b - n + n\ln(n/b)\right]&\text{for $5\sigma$ Discovery}
    \end{cases}
    \label{eq:DF_Stats}
\end{equation}

Estimation of signal yield is achieved by scaling cross section against coupling constants and multiplied by target luminosity $L=3~\si{ab ^{-1}}$. With $\sqrt{s}=3~\si{TeV}$ and $\lambda_{s\mu}=\lambda_{b\mu}=0.1$, the BSM signal cross section calculated using MG5aMC is $2.884~\si{fb}$ when LQ mass is $1~\si{TeV}$.
For background estimation, we always use the SM yields as a good approximation. The SM $s$-channel cross sections are $10.03~\si{fb}$ for $b\bar{b}$ final state and $59.63~\si{fb}$ for two light jet case. Cross sections are defined within the phase space subject to kinematics region shown in Table~\ref{tab:Cuts}.

\begin{figure}[htbp]
    \centering
    {\includegraphics[width=0.8\textwidth]{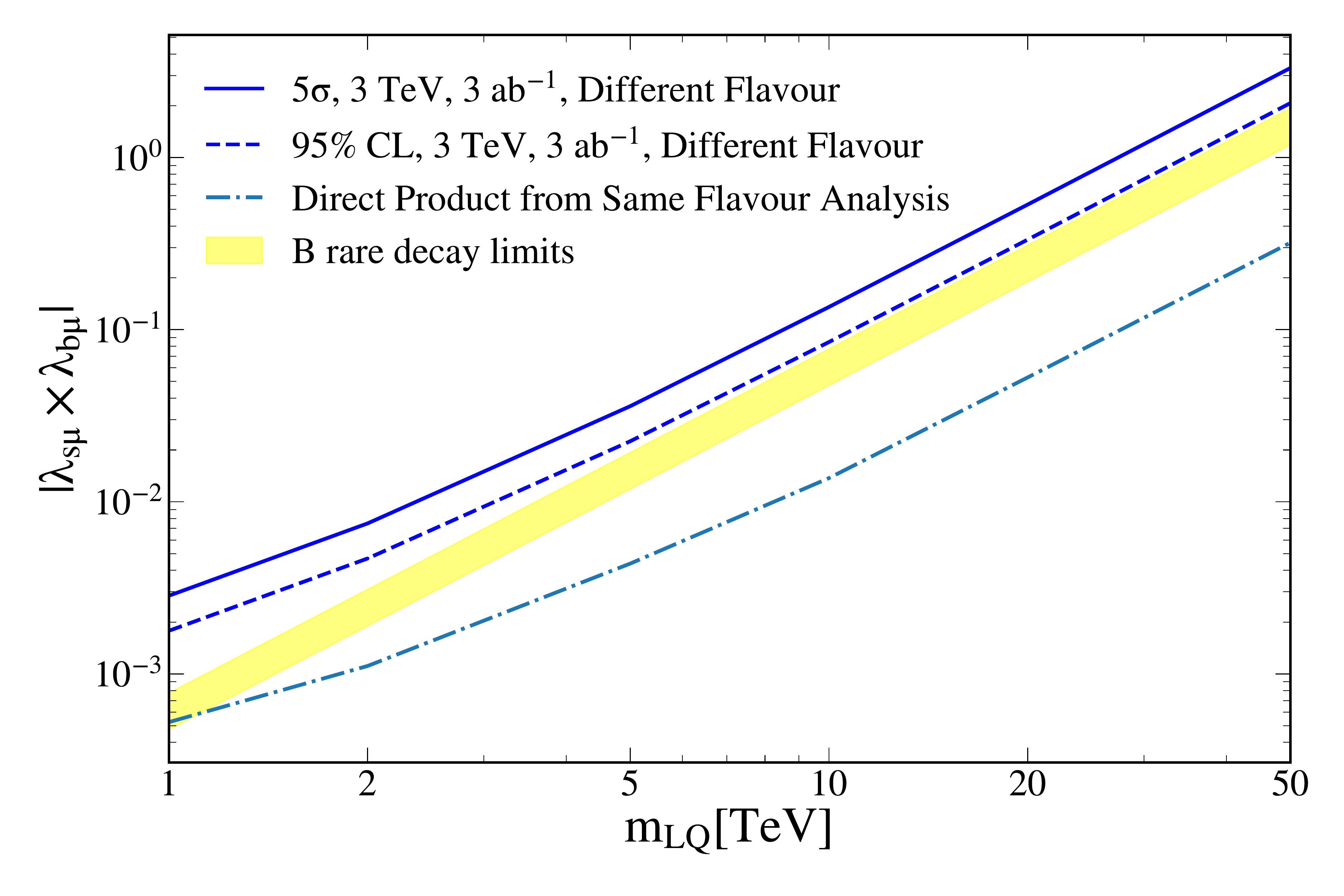}} 
    \caption{95\% CL exclusion (dashed) and 5$\sigma$ discovery limit (solid) from the different flavor jet analysis on $|\lambda_{s\mu}\bar{\lambda}_{b\mu}|$ is shown against masses of vector singlet LQ. The 2$\sigma$ parameter space favored by a fit of B anomalies~\cite{LQ1} is shown as the yellow band. Exclusion line (dash-dotted) from the same flavour analysis is also shown, by combining directly b+b and light+light results. }
    \label{fig:DF_Res}
\end{figure}

The 95\% CL exclusion line as well as 5$\sigma$ discovery limit for various LQ mass are plotted in Fig.~\ref{fig:DF_Res}. We can find that for a muon collider with $3~\si{TeV}$ C.O.M and $3~\si{ab^{-1}}$ integrated luminosity, such DF final state can approach the phase space allowed by the $B$ meson rare decay anomaly, even exclude some region with heavy LQ case. The aqua dash-dotted line is obtained via multiplying the corresponding value from the SF results, which exclude all the allowed phase space needed for the $B$ anomaly in a large range of LQ mass.


\subsection{Di-jet+di-lepton Final State: di-bjet + $\mu$ + $\tau$ as an example}

Thinking about the recent results on muon $g-2$ and anomaly of LFU, should we regard muon as the special one among all the charged leptons, or we could also consider that the lepton flavour violations could follow the mass pattern $3^{rd}>2^{nd}>1^{st}$ generation? Moreover, we have studied LQ coupling across quark generations, should we have a similar trial on comparing different lepton generations? Triggered by this idea, we start to investigate the simultaneous study on LQ-$\tau$ and LQ-$\mu$ couplings. 

\begin{figure}
    \centering
    \subfloat[LQ Pair Production\label{sfig:Pair}]{\includegraphics[width= 0.3\textwidth]{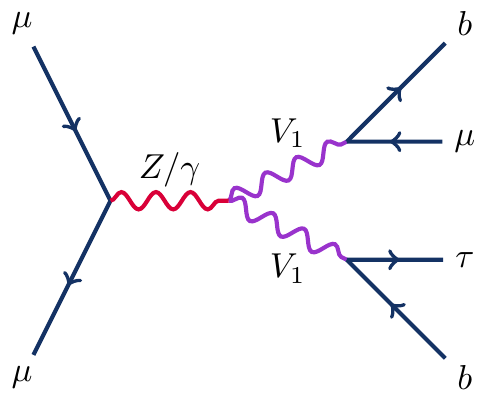}}\qquad
    \subfloat[LQ Single Production via Radiation \label{sfig:ISR_FSR}]{\includegraphics[width= 0.3\textwidth]{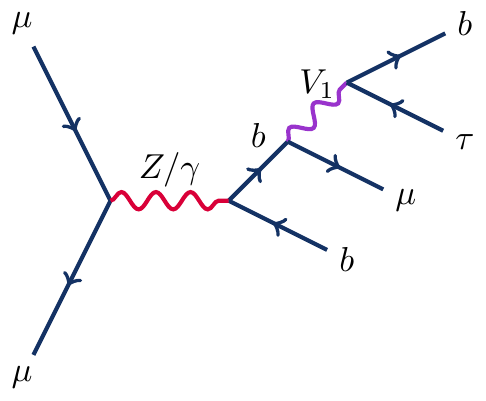}}\\
    \subfloat[LQ Single Production via Fusion\label{sfig:VXF}]{\includegraphics[width= 0.3\textwidth]{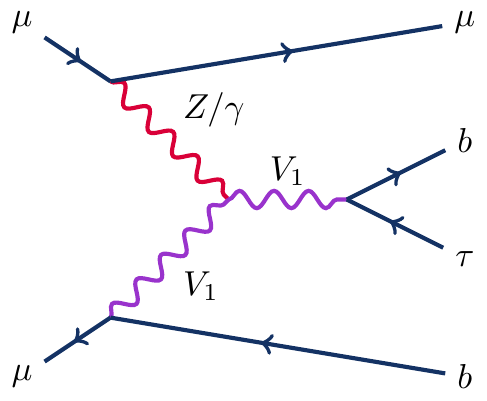}}\qquad
    \subfloat[LQ and Vector Boson Scattering \label{sfig:VXS}]{\includegraphics[width= 0.3\textwidth]{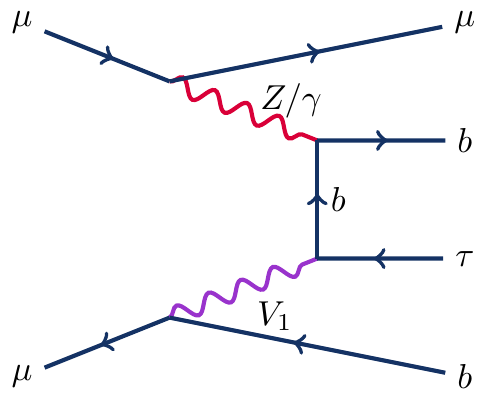}}
    \caption{LQ involved $2\to4$ processes at a muon collider, with a $\tau$ in the final states.}
    \label{fig:VXS_FD}
\end{figure}

At a muon collider, the leading contribution involving $\tau$ leptons in the final states is however not as simple as the di-jet cases. Processes including $\tau$-LQ-quark vertex can not be directly realized like the $\mu$-LQ-quark vertex which can be related to the incoming muons. Instead of a simple $2\to2$ process, we choose $2\to4$ processes containing two electroweak vertices and two LQ vertices with the order of $(\alpha_{EW}^2\cdot\lambda_{q_1l_1}\cdot\lambda_{q_2l_2})$, in which one type of process is the scattering between LQ and vector boson. Similar leptoquark involved processes with 2 jet + 2 charged lepton in the final state at muon collider is also discussed in Ref.~\cite{LQ_RK2,LQ_RK3}. However, they mainly focus on the final states with two same flavour muons (taus) and two $b$ jets, which has interference with SM processes. Since this $2\to4$ process has similar topology with vector boson scattering (VBS), we call this kind of process ``VXS''. Typical Feynman diagrams are summarized as Fig.~\ref{fig:VXS_FD}, in which both LQ pair production (Fig.~\ref{sfig:Pair}) and single production via fusion (Fig.~\ref{sfig:VXF}) contain the LQ-$Z/\gamma$-LQ vertex regarded as a electroweak vertex with order $\alpha_{EW}$. The leading contribution comes from the pair production and single production due to the $s$-channel resonance structure for a light leptoquark within the setup of interest ($\sqrt{s}\sim \mathcal{O}(\text{TeV}), m_{\text{LQ}}\sim \mathcal{O}(1\sim 10~\si{TeV})$). The ``VXS'' process will dominate the production gradually as LQ getting heavier compared with the C.O.M energy.

The final states with one $\tau$ and one $\mu$ are used in order to have the LQ-$\tau$ and LQ-$\mu$ couplings at the same time, and the jets are required to be bottom flavour (similar when requiring two same flavour light jets or two different flavour jets). The complete reconstruction of $\tau$ is not possible in the leptonic decay channel due to the existence of neutrino, so we require the $\tau$ in the final state decays hadronically, while the reconstruction of hadronic decay $\tau$ have been studied well in Ref.~\cite{tau_reco}. The hadronic decay branching fraction of $\tau$ is from PDG and is applied on the BSM signal process to estimate the yields. The cross section of BSM ($\tau\mu$bb) has a quadratic dependence on the product between $\lambda_{b\mu}$ and $\lambda_{b\tau}$ as Eq.~\ref{eq:vxs_xsec} shows. 

\begin{equation}
    \sigma(\mu^+\mu^-\to \tau\mu bb) \propto |\lambda_{b\mu}\bar{\lambda}^{b\tau}|^2
    \label{eq:vxs_xsec}
\end{equation}

Similar with different flavour di-jet final state, the di-jet+di-lepton final state have no SM background at ME level due to the violation of lepton number. However, there could be ``fake'' backgrounds from objects mis-identification, e.g., jet is mis-identificated as $\mu$ or $\tau$. We applied jet flavour tagging efficiencies as Table~\ref{tab:jet_eff} documented, and the hadronic $\tau$ identification efficiency is chosen from the performance of the CMS detector. The mis-tagging rates are both $\sim1\%$ from CEPC detector simulation~\cite{cepc_reco} and CMS estimation~\cite{tau_reco}.

\begin{figure}
    \centering

    \subfloat[Cross section under benchmark LQ coupling at 3 TeV C.O.M \label{sfig:VXS_Xsec}]{\includegraphics[width=0.4\textwidth]{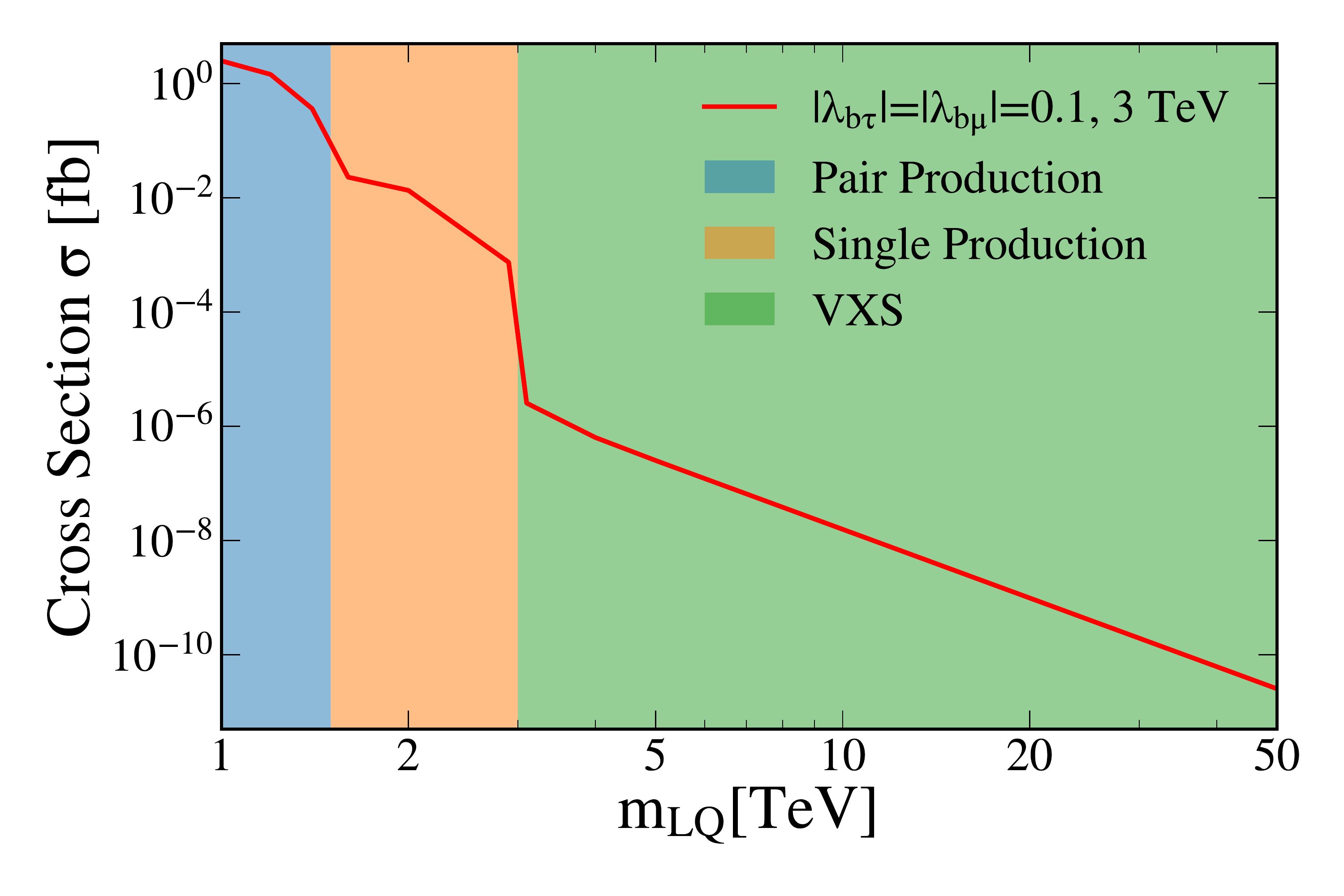}}\qquad
    \subfloat[Corresponding Limits for di-lepton + di-bjet Case \label{sfig:VXS_Res}]{\includegraphics[width=0.4\textwidth]{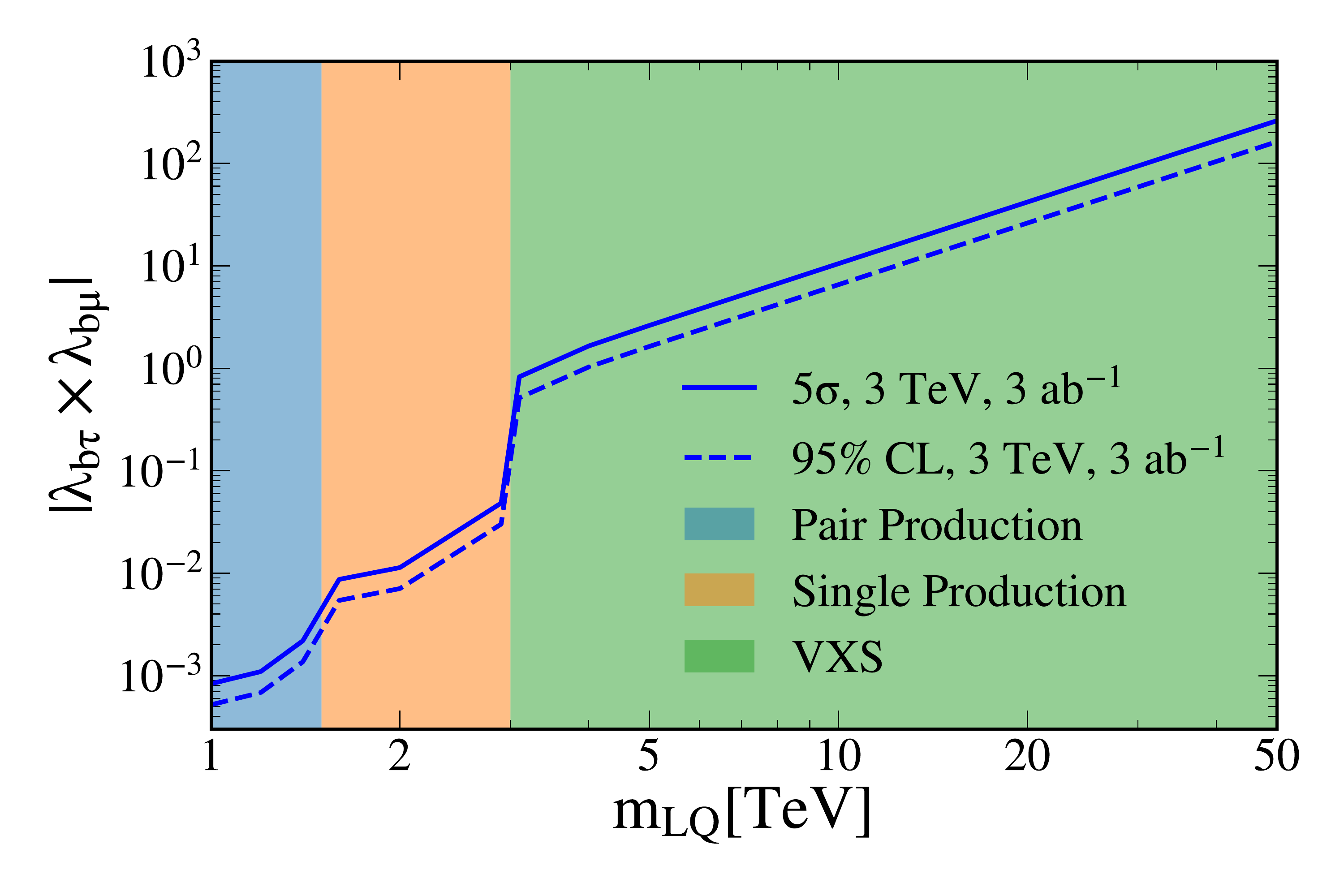}}
    \caption{(Left) Cross section dependence on LQ masses of the di-bjet and di-lepton processes, and the corresponding dominate sub-processes. (Right) 95\% CL exclusion line (dashed) and 5$\sigma$ discovery limit (solid) against LQ masses.}
    \label{fig:VXS_ALL}
\end{figure}

We use $\lambda_{b\mu}=\lambda_{b\tau}=0.1$ and $m_{LQ}=1~\si{TeV}$ as the benchmark setup, the corresponding cross section is $2.504~\si{fb}$ for $\sqrt{s}=3~\si{TeV}$ within the phase space shown in Table~\ref{tab:Cuts}. With the similar counting analysis strategy introduced for the different flavour di-jet cases, we construct $Z$ statistics as defined in Eq.~\ref{eq:DF_Stats}. The 95\% CL exclusion line as well as 5$\sigma$ discovery limit are reported in Fig.~\ref{sfig:VXS_Res}.

\section{Conclusion}

In this paper, we show the possibility to search for heavy leptoquarks at a muon collider through three channels: 1) same flavour di-jet, 2) different flavour di-jet and 3) di-jet+di-lepton final states. The leptoquark we focus on is a vector singlet, which is one kind of leptoquarks could explain the muon (g-2) anomaly and $B$ meson rare decay anomaly. We neglect the coupling between leptoquark and electron, and require that leptoquarks can only couple to down-type quarks. Under the benchmark conditions that the center-of-mass at 3 TeV and integrated luminosity $3~\si{ab^{-1}}$, we obtain the limits on $\lambda_{b\mu}$, $\lambda_{s\mu}$ and combined limit of them at different mass of leptoquark. The results show that a muon collider under the benchmark condition is capable of answering the question from $B$ meson rare decay anomaly. Besides the study on the coupling between leptoquark and $\mu$, we also show the ability of a muon collider to investigate the coupling between leptoquark and $\tau$, which enables us to understand whether the possible lepton flavour universality violation follow the mass pattern of charged leptons. 

\acknowledgments

This work is supported in part by the National Natural Science Foundation of China under Grant No. 12075004, by MOST under grant No. 2018YFA0403900.


\bibliographystyle{ieeetr}
\bibliography{h}
\end{document}